\documentclass[hidelinks,onefignum,onetabnum]{siamart251104}

\usepackage{lipsum}
\usepackage{amsfonts}
\usepackage{graphicx}
\usepackage{epstopdf}
\usepackage{algorithmic}
\ifpdf
  \DeclareGraphicsExtensions{.eps,.pdf,.png,.jpg}
\else
  \DeclareGraphicsExtensions{.eps}
\fi

\newsiamremark{remark}{Remark}
\newsiamremark{hypothesis}{Hypothesis}
\crefname{hypothesis}{Hypothesis}{Hypotheses}
\newsiamthm{claim}{Claim}
\newsiamremark{fact}{Fact}
\crefname{fact}{Fact}{Facts}

\headers{The stability analysis}{Q. GAO, X. JI,  Z. DU, S. LI, Y. CHEN, and K. XU}

\title{The stability priority of spatial-temporal coupled compact element methods over decoupled compact element methods\thanks{Submitted to the editors 
\funding{This work was funded by the Funding of National Key Laboratory of Computational Physics, the National Natural Science Foundation of China (Grant No. 12302378, 12172316, and 92371107) and the Natural Science Basic Research Plan in Shaanxi Province of China (No. 2025SYS-SYSZD-070).}}}

\author{Qihui GAO\thanks{State Key Laboratory for Strength and Vibration of Mechanical Structures, Shaanxi Key Laboratory of Environment and Control for Flight Vehicle, School of Aerospace Engineering, Xi’an Jiaotong University, Xi’an, China 
  (\email{qhgao@stu.xjtu.edu.cn}, \email{jixing@xjtu.edu.cn}).}
\and Xing JI\footnotemark[2]
\and ZHIFANG DU\thanks{Institute of Applied Physics and Computational Mathematics, Beijing, PR China (\email{du$\_$zhifang@iapcm.ac.cn}, \email{lishiyi14@tsinghua.org.cn }, \email{chen$\_$yibing@iapcm.ac.cn})}
\and  Shiyi LI \footnotemark[3]
\and  Yibing CHEN \footnotemark[3]
\and Kun XU\thanks{{Department of Mathematics, Hong Kong University of Science and Technology, Clear Water Bay, Kowloon, Hong Kong (\email{makxu@ust.hk})}, {Department of Mechanical and Aerospace Engineering, Hong Kong University of Science and Technology, Clear Water Bay, Kowloon, Hong Kong}, and {Shenzhen Research Institute, Hong Kong University of Science and Technology, Shenzhen, China}}
}

\usepackage{amsopn}

\ifpdf
\hypersetup{
  pdftitle={The stability priority of spatial-temporal coupled compact element methods over decoupled compact element methods},
  pdfauthor={Q. GAO, X. JI,  Z. DU, S. LI, Y. CHEN, and K. XU}
}
\fi

\begin{document}

\maketitle

\begin{abstract}
With the increasing industrial demands, two families of high-order numerical schemes are widely used within the computational fluid dynamics community. One is the method of line, which relies on Runge-Kutta (RK) time-stepping applied to a semi-discrete, spatio-temporally decoupled formulation. The other is the family of Lax-Wendroff (LW) type method, which are inherently spatial-temporal coupled and are constructed within a multi-stage multi-derivative (MSMD) framework. This paper, for the first time, conducted a comparative Fourier stability analysis of RK and LW method to distinguish the dispersion and dissipation effects of numerical schemes respectively. Through rigorous theoretical derivation and consistent numerical validation, we draw the following conclusions: While explicit RK line methods are straightforward like Discontinuous Galerkin (DG) method and flux reconstruction (FR) method, they employ from a decoupling of spatial and temporal accuracy, thus discarding flow field evolution information and requiring small time steps. In contrast, spatial-temporal coupled compact methods, such as the gas-kinetic scheme (GKS) and the generalized Riemann problem (GRP) solver, utilize initial-value information from space far more effectively for time evolution. Even with just one additional order of spatial-temporal coupled information, they show better stability compared to RK methods. This provides new insights for CFD algorithm design, emphasizing the need for consistency between the dependence in the physical domain and  that in numerical domain.
\end{abstract}




\section{Introduction}
 Over past decades, the computational fluid dynamics field has witnessed significant progress in high-order numerical methods for hyperbolic conservation laws. In the simulation of compressible fluid flows or associated problems, two families of space-time discretization methods are commonly-used. One family is the methods of line, i.e. spatial-temporal decoupled method, for which the fluid dynamical system is written in a semi-discrete form and the Runge-Kutta (RK) temporal iteration is then employed for time marching. The Runge-Kutta scheme use a multi-stage procedure to achieve high-order accuracy, featuring very favorable properties such as simplicity in time-stepping. Notable advancements include, such as RK-WENO \cite{JS,QS1}, RK-DG \cite{BS1}, RK-FR \cite{Huynh} and their variants, etc, demonstrating great potential for complex engineering problems. The RK-DG method \cite{BS1,BS2} combines the approximation flexibility of the finite element method with Runge-Kutta temporal discretization in finite volume methods. It maintains compact stencil requirements while achieving high-order spatial accuracy. DG methods demonstrate remarkable strengths within mathematical frameworks and exhibit  promising prospects for simulating convection dominated problems, such as linear and nonlinear waves including shock waves. Nevertheless, they seem to lack robustness when the CFL number is restricted by the order of polynomial. Still, the development of DG methods persists as a main research direction in high-order numerical schemes. The RK-FR method \cite{Huynh} bridges the discontinuous Galerkin and spectral difference,resulting in simplified versions of both. The FR method, a class of discontinuous spectral element methods for discretizing conservation laws, utilizes a nodal basis defined on some solution points like Gauss points for approximating the solution with piecewise polynomials. The main idea involves constructing a continuous flux approximation through using a correction function and numerical flux at the cell interfaces. The nodal solution values are then updated via a collocation scheme coupled with a Runge-Kutta method. Runge-Kutta schemes use multi-stage procedure to achieve the high order, which, due to the CFL condition $\Delta t=O(\Delta x)$, requires at least as many stages as the formal order of accuracy. The need to conduct multi-stage of RK scheme nature implies that costly limiters and data exchange in parallel computations must be performed multiple times per time step \cite{DFTB}. Furthermore, some order barriers exist at high orders, where even more stages than the order are often required. 
 
 In contrast to RK scheme, the other major family is the Lax-Wendroff (LW) type method, which achieves high order temporal accuracy within the multi-stage multi-derivative (MSMD) framework by combining the simplicity of multi-stage RK methods with the spatial-temporal coupling inherent to one-stage LW type methods \cite{LD}. The MSMD method achieves success in solvers such as the generalized Riemann problem (GRP) and the gas kinetic scheme (GKS), enabling the construction of high order spatial-temporally coupled schemes with computational performance. The GRP solver extends the first order Godunov solver to second order temporal accuracy using MUSCL type initial data \cite{V}. Its acoustic version, known as the arbitrary derivative (ADER) solver \cite{TT,TT2}, can also serve as a building block. They implemented finite-volume ADER schemes featuring the new Derivative Riemann solver for the compressible Euler equations, achieving very high-order accuracy and essentially non-oscillatory solutions \cite{TT}. Another alternative is the gas kinetic scheme (GKS) \cite{X2,XH}, which has been systematically developed and validated for simulating Euler and Navier-Stokes (N-S) flows based on mesoscopic gas kinetic theory. GKS achieves high precision in smooth regions while maintaining robustness near discontinuities \cite{JPSX}. Within the MSMD framework for ODEs \cite{HW}, a two-stage fourth-order GKS with a second-order flux function was established for N-S equations \cite{PXL,PJCWX}. Subsequent extensions integrated second-order GKS fluxes with compact spatial operators, the Compact Gas-Kinetic Scheme (CGKS) has been optimized. These innovations equip CGKS with a strong capabilities in resolving complex shock interactions while preserving shock-capturing stability, showcasing high computational efficiency within a physics-consistent framework through high-order gas-kinetic evolution. Numerical experiments with both GRP and GKS solvers confirm their suitability for designing such fourth-order accurate method \cite{LD,PXL}.
 
 When achieving fourth-order temporal accuracy, RK-type compact element methods and LW-type ones exhibit both shared traits and fundamental distinctions in computational performance. This can be observed by comparing representative solvers such as DG and GKS. Both methods employ compact stencils and HWENO reconstruction for spatial discretization to compute numerical fluxes and apply monotonicity limiters. The DG method attains fourth-order accuracy through a four-stage RK scheme, requiring four intermediate steps per time step with CFL stability limit restricted to $1/9$. In contrast, the compact GKS adopts the two-stage fourth-order method, raising the permissible CFL number to 0.5 while halving computational costs. The DG method typically relies on a weak formulation within a spatial-temporal decoupled Runge-Kutta framework. In contrast, the compact GKS directly updates the slopes using explicitly evolved interface values, achieving a fully coupled discretization and thereby embedding  consistent numerical dissipation. Notably, the fourth-order compact GKS implements HWENO reconstruction without requiring any additional trouble-cell detector\cite{V}. These characteristics stem from the inherent gas-kinetic model, which intrinsically unifies microscopic particle dynamics with macroscopic conservation laws. Unlike the purely mathematical construction of DG, GKS incorporates physical relaxation processes through its kinetic theory, enabling it to maintain stability at much larger time steps compared to the same-order DG method. Recently, Mu and Ji developed a very high order spatial reconstruction for CGKS within CFL number at 0.5 for strong compressible flow, which exhibit excellent robustness in capturing discontinuous and complex flow structures \cite{MZJ}.
 
 Although GKS and CGKS have shown considerable potential and broad applicability across various numerical simulations, their linear stability analysis remains less extensively developed compared to DG and FR methods, as the latter benefits from systematic theoretical frameworks \cite{BS1,BS2,Huynh}. Xie et al. \cite{XJX} analyzed a stability analysis for the spectral difference gas-kinetic scheme, revealing that GKS achieves higher efficiency with fewer stages despite having slightly smaller permissible CFL numbers than conventional methods. Beyond this, theoretical studies on the stability of GKS remain scarce, with most research focusing on numerical fluid dynamics simulations. As previously noted, the pronounced contrast between DG method and CGKS under fourth-order temporal discretization underscores the necessity of in-depth stability analysis investigations for CGKS.
 
 Li and Li \cite{LL} proposed a hybrid reconstruction approach to balance the efficiency of second-order $(k = 1)$ and higher-order $(k\geq2)$ Hermite reconstruction within a linear stability analysis framework. By integrating Hermite-type reconstructions with Lax–Wendroff-type time discretizations, they developed a compact space-time coupled scheme. Their study treats both the solution itself $\bar{u}_{j}^{n+1}$ and its derivative average ${D u}_{j}^{n+1}$ in a multi-moment schemes, which implements linear dissipation analysis within this unified formulation.
 
 For large-time-step simulations of convection-dominated nonlinear problems, the studies above  establish that adequate linear stability is an essential criterion when selecting numerical schemes. In this paper, the Fourier analysis is conducted to compare dissipation behaviors between RK and LW-type compact element methods under identical second-order two-moment schemes systematically, offering theoretical guidance for optimal scheme selection. Within the two-moment Fourier modal framework, the differing dissipative properties of DG and CGKS are quantified through eigenvalue distributions in the complex plane, which reveals key differences in the spectral properties of their respective numerical dissipation mechanisms.
 
 This paper is organized as follows. Section \ref{sec2} provides a detailed description of the representative RK and LW-type solvers. Section \ref{sec3} presents the dissipation properties and stability of the two-moment schemes, extending the stability analysis for high-order CGKS. Finally, Section \ref{sec4} discusses the advantages of the spatial-temporal coupled method over decoupled ones from a multi-physics perspective.

\section{Basic framework of compact element methods: LW type and RK type}\label{sec2}
\subsection{GRP framework}
Considering hyperbolic conservation laws 
\begin{equation}\label{eq1.1}
  \frac{\partial \boldsymbol{u}}{\partial t}+\frac{\partial \boldsymbol{f}(\boldsymbol{u})}{\partial x}=0, x \in \mathbb{R}, t>0, 
\end{equation}
where the initial value satisfies $\boldsymbol{u}(x, 0)=\boldsymbol{u}_{0}(x), x \in \mathbb{R}$ and $\boldsymbol{f}(\boldsymbol{u})$ is the associated flux function vector. Given the computational mesh $I_{j}=(x_{j-\frac{1}{2}}, x_{j+\frac{1}{2}})$ with the size $h=x_{j+\frac{1}{2}}-x_{j-\frac{1}{2}}$ for every $j$, write \eqref{eq1.1} in form of the balance law,
\begin{equation*}
\frac{d \overline{\boldsymbol{u}}_{j}(t)}{d t}=\mathcal{L}_{j}(\boldsymbol{u}):=-\frac{1}{h}\left[\boldsymbol{f}(\boldsymbol{u}(x_{j+\frac{1}{2}}, t))-\boldsymbol{f}(\boldsymbol{u}(x_{j-\frac{1}{2}}, t))\right], \quad \overline{\boldsymbol{u}}_{j}(t)=\frac{1}{h} \int_{I_{j}} \boldsymbol{u}(x, t) dx
\end{equation*}
where $\mathbf{u}(x_{j+\frac{1}{2}}, t)$ is described using the following GRP solver \cite{BF}.

Given the piecewise-linear distribution $\mathbf{u}(x, t^{n})=\mathbf{u}^{n}(x)$ by the HWENO interpolation, compute the corresponding GRP value $\boldsymbol{u}_{j+\frac{1}{2}}^{n},(\partial \boldsymbol{u} / \partial t)_{j+\frac{1}{2}}^{n}$ and construct $\overline{\boldsymbol{u}}^{n+1}(x)$ to approximate the exact solution $\tilde{\boldsymbol{u}}\left(x, t_{n}+k\right)$ ) as follows.

Step 1. At every cell boundary $x_{j+1 / 2}$ evaluate interface values $\mathbf{u}_{j+\frac{1}{2}}^{n}$ with the cell averages $\overline{\mathbf{u}}_{j}^{n}$
\begin{equation*}
\boldsymbol{u}_{j+1 / 2, \pm}^{n}=\lim _{\delta \rightarrow 0^{+}} \boldsymbol{u}^{n}\left(x_{j+1 / 2} \pm \delta\right)= \begin{cases}\overline{\boldsymbol{u}}_{j+1}^{n}-\frac{\Delta x}{2} s_{j+1}^{n}, & +\\ \overline{\boldsymbol{u}}_{j}^{n}+\frac{\Delta x}{2} s_{j}^{n}, &-.\end{cases}    
\end{equation*}

Then determine the Riemann solution

\begin{equation*}
\boldsymbol{u}_{j+1 / 2}^{n}=R(0 ; \boldsymbol{u}_{j+1 / 2,-}^{n}, \boldsymbol{u}_{j+1 / 2,+}^{n}),
\end{equation*}
that is
\begin{align*}
\boldsymbol{u}_{j+1 / 2}^{n}= \begin{cases}\boldsymbol{u}_{j+1 / 2,-}^{n} & \text { wave moves right, } f^{\prime}(\boldsymbol{u}_{j+1 / 2}^{n})>0,  \\ \boldsymbol{u}_{j+1 / 2,+}^{n} & \text { wave moves left, } f^{\prime}(\boldsymbol{u}_{j+1 / 2}^{n})<0, \\ \boldsymbol{u}_{\min } & \text { if } x_{j+1 / 2} \text { is a sonic point, } f^{\prime}(\boldsymbol{u}_{j+1 / 2,-}^{n}) \leq 0 \leq f^{\prime}(\boldsymbol{u}_{j+1 / 2,+}^{n}) .\end{cases}  
\end{align*}

Step 2. Determine the instantaneous time derivatives $\frac{\partial \tilde{\mathbf{u}}}{\partial t}(x_{j+1 / 2}, t_{n})$ by
\begin{align*}
\frac{\partial \tilde{\boldsymbol{u}}}{\partial t}\left(x_{j+1 / 2}, t_{n}\right)= \begin{cases}-f^{\prime}(\boldsymbol{u}_{j+1 / 2}^{n}) s_{j}^{n} & \text { if } f^{\prime}(\boldsymbol{u}_{j+1 / 2}^{n})>0  \\ -f^{\prime}(\boldsymbol{u}_{j+1 / 2}^{n}) s_{j+1}^{n} & \text { if } f^{\prime}(\boldsymbol{u}_{j+1 / 2}^{n})<0 \\ 0 & \text { if } \boldsymbol{u}_{j+1 / 2}^{n}=\boldsymbol{u}_{\mathrm{min}}\end{cases}
\end{align*}

Then compute the approximate solution and numerical flux at the midpoint $\left(x_{j+1 / 2}, t_{n+1 / 2}\right)$ by

\begin{equation*}
\boldsymbol{u}_{j+1 / 2}^{n+1 / 2}  =\boldsymbol{u}_{j+1 / 2}^{n}+\frac{k}{2} \frac{\partial \tilde{\boldsymbol{u}}}{\partial t}(x_{j+1 / 2}, t_{n}) 
\end{equation*}
\begin{equation*}
    f_{j+1 / 2}^{n+1 / 2}  =f(\boldsymbol{u}_{j+1 / 2}^{n})+\frac{k}{2} f^{\prime}(\boldsymbol{u}_{j+1 / 2}^{n}) \frac{\partial \tilde{\boldsymbol{u}}}{\partial t}(x_{j+1 / 2}, t_{n})
\end{equation*}

Step 3. Evaluate the new cell averages
\begin{equation*}
\overline{\boldsymbol{u}}_{j}^{n+1}=\overline{\boldsymbol{u}}_{j}^{n}-\lambda\left(\boldsymbol{f}_{j+1 / 2}^{n+1 / 2}-\boldsymbol{f}_{j-1 / 2}^{n+1 / 2}\right), \quad-\infty<j<\infty, 
\end{equation*}
and the new slopes by
\begin{align*}
\boldsymbol{u}_{j+1 / 2}^{n+1} & =\boldsymbol{u}_{j+1 / 2}^{n}+k \frac{\partial \tilde{\boldsymbol{u}}}{\partial t}(x_{j+1 / 2}, t_{n}), \quad-\infty<j<\infty  \\
s_{j}^{n+1} & =\frac{1}{\Delta x}(\boldsymbol{u}_{j+1 / 2}^{n+1}-\boldsymbol{u}_{j-1 / 2}^{n+1})
\end{align*}

\subsection{CGKS framework}
The gas-kinetic BGK equation can be written as
\begin{equation}\label{eq2.1}
\frac{\partial f}{\partial t}+u_{i} \frac{\partial f}{\partial x_{i}}=\frac{g-f}{\tau}
\end{equation}
where $f$ is the gas distribution function, $g$ is the corresponding equilibrium state, and $\tau$ is the collision time. Due to the mass, momentum and energy conservation in particle collisions, the collision term satisfies the following compatibility condition,
\begin{equation}
\int \frac{g-f}{\tau} \psi \mathrm{~d} \Xi=0 
\end{equation}
where $\psi=\left(1, u, \frac{1}{2}\left(u^{2}+\xi^{2}\right)\right)^{T}, \mathrm{~d} \Xi=\mathrm{d} u \mathrm{~d} \xi_{1} \ldots \mathrm{~d} \xi_{K}$, the variable $\xi^{2}=\xi_{1}^{2}+\xi_{2}^{2}+\ldots+\xi_{K}^{2}$. The above BGK model coincides in form with the equations in the theory of relaxation processes and if $\tau$ is a local constant, the gas distribution function could be written in the following integral form
\begin{equation}\label{eq2.3}
f\left(x_{i}, t, u_{i}, \xi\right)=\frac{1}{\tau} \int_{0}^{t} g\left(x_{i}-u_{i}(t-t^{\prime}), t^{\prime}, u_{i}, \xi\right) e^{-\left(t-t^{\prime}\right) / \tau} d t^{\prime}+e^{-t / \tau} f_{0}(x_{i}-u_{i}(t-t_{0}),t_{0}, u_{i}, \xi)
\end{equation}
where $f_{0}$ is the real gas distribution function $f$ at $t_{0}$, and $g$ is the corresponding equilibrium state. The integral solution essentially describes a physical process from the particle-free transport governed by $f_{0}$ on the kinetic scale to the evolution of the macroscopic hydrodynamic flow in the integral term $g$. 

The equivalent integral solution of $f$ at the cell interface and time $t$ is
\begin{align*}
f(x, t, u_{i}, \xi)= & \left(1-e^{-t / \tau_{n}}\right) g^{c}+\left((t+\tau) e^{-t / \tau_{n}}-\tau\right) a^{c} u g^{c}+\left(t-\tau+\tau e^{-t / \tau_{n}}\right) A^{c} g^{c} \\
& +e^{-t / \tau_{n}} g^{l / r}\left[1-(t+\tau) a^{l / r} u_{i}-\tau A^{l / r}\right]
\end{align*}
where $a^{c}=g_{x}^{c} / g^{c}$, $a^{l / r}=g_{x}^{l / r} / g^{l / r}$, $A^{c}=g_{t}^{c} / g^{c}$, $A^{l / r}= g_{t}^{l / r} / g^{l / r}$ and the coefficients $a^{k}, A^{k}, k=l, r$ are defined according to the expansion of $g_{k}$. $\tau$ is enlarged as
\begin{equation*}
\tau_{n}=\tau+c_{2}\left|\frac{p_{l}-p_{r}}{p_{l}+p_{r}}\right| \Delta t=\frac{\mu}{p}+c_{2}\left|\frac{p_{l}-p_{r}}{p_{l}+p_{r}}\right| \Delta t.   
\end{equation*}
Besides, $g^{c}$ is obtained by
\begin{equation*}
\int_{-\infty}^{\infty} d \xi \int_{-\infty}^{\infty} \psi_{\alpha} g^{c} d u=\int_{-\infty}^{\infty} d \xi \int_{0}^{\infty} \psi_{\alpha} g^{l} d u+\int_{-\infty}^{\infty} d \xi \int_{-\infty}^{0} \psi_{\alpha} g^{r} d u.
\end{equation*}

By selecting the appropriate equilibrium state in the BGK model, most well-known viscous conservation laws can be recovered to a certain degree. The inviscid hyperbolic system corresponds to the state with a local equilibrium distribution function.

Considering the one-dimensional linear advection-diffusion equation
\begin{equation}
u_{t}+c u_{x}=\mu u_{x x},
\end{equation}
where $\mu$ is the viscosity coefficient. The above equation can be derived from the BGK model by adopting the equilibrium state \cite{X2}
\begin{equation}
  g=U\left(\frac{\lambda}{\pi}\right)^{1 / 2} e^{-\lambda(u-c)^{2}},
\end{equation}
and satisfies the conservation constraint
\begin{equation}
 \int_{-\infty}^{\infty}(f-g) d u=0,
\end{equation}

From the Chapman-Enskog expansion, the first order 
truncated distribution function $f$ is
\begin{equation}
f=g-\tau\left(g_{t}+u g_{x}\right).
\end{equation}

Substituting the above equation into the BGK model and integrating with respect to $u$, the linear advection equation can be recovered. Using the  equation \eqref{eq2.3}, the gas distribution function at the cell interface satisfies  
\begin{equation}
    f(x_{i+\frac{1}{2}},t,u,\xi)=\frac{1}{\tau}\int_{0}^{t}g(x',t',u,\xi)e^{(t-t')/\tau}dt'+e^{-t/\tau}f_{0}(-ut,u,\xi),
\end{equation}
where $x_{i+\frac{1}{2}}=0$ is the location of cell interface and $x_{i+1 / 2}=x^{\prime}+u\left(t-t^{\prime}\right)$ is the trajectory of particles. Neglect the viscous contribution, i.e. $\tau=0$, and the time dependent gas distribution function $f_{i+\frac{1}{2}}$ at the cell interface $x_{i+\frac{1}{2}}$ for \eqref{eq2.1} goes to
\begin{equation}
 f(x_{i+\frac{1}{2}},t,u,\xi)=g_{0}+t\bar{A}g_{0},
\end{equation}
where $\bar{A}=\frac{\partial g}{\partial t}/g_{0}$.

Therefore, the macroscopic variable can be obtained by integrate the Maxwell distribution function on the velocity space
\begin{equation}
    F_{i+\frac{1}{2}}=\int u\psi f_{i+\frac{1}{2}}d\Xi,
\end{equation}
where $u=c$, $\psi=1$ for \eqref{eq2.1}.
For the gas-kinetic scheme, the gas evolution is a relaxation process from kinetic to hydrodynamic scale through the exponential function, and the corresponding flux is a complicit function of time to obtain. Apply the gas-kinetic flux in the cell interfaces, then the one stage second-order method (S1O2) or second-order Runge-Kutta method (RK2) are employed for temporal discretization, which will be shown in section \ref{sec3}.

\subsection{DG framework}
Considering one-dimensional hyperbolic conservation laws \eqref{eq1.1}, where the initial value satisfies $\boldsymbol{u}(x, 0)=\boldsymbol{u}_{0}(x), x \in \mathbb{R}$, $\boldsymbol{u}$ is a vector of conservative variables and $\boldsymbol{f}(\boldsymbol{u})$ is the associated flux function vector. To solve \eqref{eq1.1} with discontinuous Galerkin method, the first step is to give a partition of the computational domain, including cells 
\begin{equation}
I_j = \left[ x_{j-\frac{1}{2}}, x_{j+\frac{1}{2}} \right], \quad j = 1, \cdots, N
\end{equation}
where the cell center is denoted by
\begin{equation}
x_j = \frac{1}{2} \left( x_{j-\frac{1}{2}} + x_{j+\frac{1}{2}} \right)
\end{equation}
and the mesh size by
\begin{equation}
\Delta x_j = x_{j+\frac{1}{2}} - x_{j-\frac{1}{2}}
\end{equation}

The solution, as well as the test function space for the Galerkin method, is given by
\begin{equation}
V_h^k = \{ p : p|_{I_j} \in P^k(I_j) \}
\end{equation}
where $P^k(I_j)$ is the space of polynomials of degree $\leq k$ on the cell $I_j$. In this paper, a Legendre polynomials
\begin{equation}
\left\{ v_l^{(j)}(x) \right\}_{l=0}^k,
\end{equation}
serving as an orthogonal basis over $I_j$, is utilized to avoid ill-conditioning of the mass matrix and $k=2$ in the following derivation.

The one-dimensional solution $u_h(x,t) \in V_h^k$ can be expressed as:
\begin{equation}\label{eq2.16}
u_h(x,t) = \sum_{l=0}^{k} u_j^{(l)}(t) v_l^{(j)}(x), \quad x \in I_j
\end{equation}

The semi-discrete scheme seeks a unique function $u_h(\cdot,t) \in V_h^k$ satisfying:
\begin{equation}\label{eq:semidiscrete}
\int_{I_j} \frac{\partial u_h}{\partial t} v \,dx - \int_{I_j} f(u_h) \frac{\partial v}{\partial x} \,dx 
+ \hat{f}_{j+\frac{1}{2}} v(x_{j+\frac{1}{2}}^-) 
- \hat{f}_{j-\frac{1}{2}} v(x_{j-\frac{1}{2}}^+) = 0
\end{equation}
 for all test functions $v \in V_h^k$. The upwind flux is adopted for numerical stabilization at cell interfaces accordingly.

 Simplify the above equation to get
\begin{equation}\label{eq2.18}
\frac{\mathrm{d}}{\mathrm{d}t} \int_{I_j} u_{h}v \,dx - \int_{I_j} f(u_h) \frac{\partial v}{\partial x} \,dx 
+ \hat{f}_{j+\frac{1}{2}} v(x_{j+\frac{1}{2}}^-) 
- \hat{f}_{j-\frac{1}{2}} v(x_{j-\frac{1}{2}}^+) = 0
\end{equation}

 The spatial discretization yields a system of ordinary differential equations: 
\begin{equation}\label{eq2.19}
\frac{\mathrm{d}\boldsymbol{u}_{j}}{\mathrm{d}t}=\boldsymbol{M}^{-1}\boldsymbol{R}(\boldsymbol{u}_{j})
\end{equation}
where in the above formula $\boldsymbol{u}_{j}=(u_{j}^{(0)}(t),u_{j}^{(1)}(t),\dots ,u_{j}^{(k)}(t))^{\mathrm{T}}$ is the vector of unknown degrees of freedom, $\boldsymbol{M}$ denotes the mass matrix,
and $\boldsymbol{R}\left(\boldsymbol{u}_{j}\right)$ is the vector obtained from spatial discretization \eqref{eq2.18}.
 In order to achieve the corresponding order temporal accuracy, the explicit Runge–Kutta scheme is adopted to solve the ordinary equation system. Furthermore, the general formulation of the one stage second-order scheme for solving the above ODEs is employed for temporal discretization, which can be given in Section \ref{sec3}.
\subsection{FR framework}
 Consider scalar conservation law of the form \eqref{eq1.1}, $u$ is conserved quantity, $f(u)$ is the corresponding flux, together with some initial and
boundary conditions. Divide the computational domain $\Omega$ into disjoint elements $\Omega_{e}=[x_{e
\frac{1}{2}}, x_{e+\frac{1}{2}}]$ and map each element to a reference element, $\Omega_{e}
\rightarrow[0,1]$, by
\begin{equation*}
    x \rightarrow \xi=\frac{x-x_{e-\frac{1}{2}}}{\Delta x_{e}},
\end{equation*}
with $\Delta x_{e}=x_{e+\frac{1}{2}}-x_{e-\frac{1}{2}}$. 
 
To approximate the solution by degree $N \geq 0$ polynomials belonging to the set $\mathbb{P}_{N}$. For this, choose $N+1$ distinct nodes
\begin{equation*}
    0 \leq \xi_{0}<\xi_{1}<\cdots<\xi_{N} \leq 1
\end{equation*}
which will be taken to be Gauss-Legendre (GL) or Gauss-Lobatto-Legendre (GLL) nodes, and will also be referred to as solution points.The solution inside an element is given by
\begin{equation*}
    x \in \Omega_{e}: \quad u_{h}(\xi, t)=\sum_{j=0}^{N} u_{j}^{e}(t) \ell_{j}(\xi)
\end{equation*}
where each $\ell_{j}$ is a Lagrange polynomial of degree $N$. Note that the coefficients $u_{j}^{e}$ which are the basic unknowns or degrees of freedom, are the solution values at the solution points. 
The FR scheme is based on spatial discretization to construct a continuous polynomial approximation of the flux which is then used in a collocation scheme to update the nodal solution values. At some time $t$, the FR scheme can be described by the following steps.

Step 1. In each element, construct the flux approximation by interpolating the flux at the solution points leading to a polynomial of degree $N$, given by
\begin{equation}\label{eq2.21}
f_{h}^{\delta}(\xi, t)=\sum_{j=0}^{N} f\left(u_{j}^{e}(t)\right) \ell_{j}(\xi)
\end{equation}

Step 2. Build a continuous flux approximation by adding some correction terms at the element boundaries
\begin{equation}\label{eq2.22}
f_{h}(\xi, t)=\left[f_{e-\frac{1}{2}}(t)-f_{h}^{\delta}(0, t)\right] g_{L}(\xi)+f_{h}^{\delta}(\xi, t)+\left[f_{e+\frac{1}{2}}(t)-f_{h}^{\delta}(1, t)\right] g_{R}(\xi)   
\end{equation}
where$f_{e+\frac{1}{2}}(t)=f(u_{h}(x_{e+\frac{1}{2}}^{-}, t), u_{h}(x_{e+\frac{1}{2}}^{+}, t))$
is a numerical flux function that makes the flux unique across the cells. The functions $g_{L}, g_{R}$ are the correction functions which must be chosen to obtain a stable scheme.

Step 3. Obtain the system of ODEs by collocating the PDE at the solution points
\begin{equation}\label{eq2.23}
\frac{\mathrm{d} u_{j}^{e}}{\mathrm{~d} t}(t)=-\frac{1}{\Delta x_{e}} \frac{\partial f_{h}}{\partial \xi}(\xi_{j}, t), \quad 0 \leq j \leq N    
\end{equation}
which is solved in time by time marching numerical schemes.

The correction functions $g_{L}, g_{R}$ should satisfy the end point conditions
\begin{align}
g_{L}(0)=1, & g_{R}(0)=0 \nonumber\\
g_{L}(1)=0, & g_{R}(1)=1\nonumber
\end{align}
which ensures the continuity of the flux, i.e., $f_{h}(x_{e+\frac{1}{2}}^{-}, t)=f_{h}(x_{e+\frac{1}{2}}^{+}, t)=f_{e+\frac{1}{2}}(t)$. There is a wide family of correction functions available in the literature \cite{Huynh}. Two of these functions, the Radau and $g_2$ correction functions, are of major interest since they correspond to commonly used DG formulations. The Radau correction function is a polynomial of degree $N+1$ which belongs to the family of \cite{Huynh} corresponding to the parameter $c=0$ and given by
\begin{equation*}
g_{L}(\xi)=\frac{(-1)^{N}}{2}\left[L_{N}(2 \xi-1)-L_{N+1}(2 \xi-1)\right], \quad g_{R}(\xi)=\frac{1}{2}\left[L_{N}(2 \xi-1)+L_{N+1}(2 \xi-1)\right]    
\end{equation*}
where $L_{N}:[-1,1] \rightarrow \mathbb{R}$ is the Legendre polynomial of degree $N$. In the general class of [74], $g_{2}$ correction function of degree $N+1$ corresponds to $c=\frac{2(N+1)}{(2 N+1) N\left(a_{N} N!\right)^{2}}$ where $a_{N}$ is the leading coefficient of $L_{N}$; they are given by
\begin{align}
& g_{L}(\xi)=\frac{(-1)^{N}}{2}\left[L_{N}(2 \xi-1)-\frac{(N+1) L_{N-1}(2 \xi-1)+N L_{N+1}(2 \xi-1)}{2 N+1}\right] \\
& g_{R}(\xi)=\frac{1}{2}\left[L_{N}(2 \xi-1)+\frac{(N+1) L_{N-1}(2 \xi-1)+N L_{N+1}(2 \xi-1)}{2 N+1}\right]
\end{align}

\begin{remark}\label{re2.1}
It can be shown that the RKFR scheme with the Radau and g2 correction functions is identical to the nodal RKDG method employing GL and GLL nodes respectively \cite{VC}.
\end{remark}

\section{The analysis of stability property of fully discretized form of two moment LW type and RK type solvers}\label{sec3}

\subsection{Dissipation analysis for second-order method}\label{sec3.4}
 Linear dissipation analysis serves as a useful instrument for probing numeriacl sheme stability. As  the aforementioned schemes belong to the multi-moment category — simultaneously evolving the solution and its spatial derivative averages — they are classified as two-moment schemes, necessitating dissipation characterization through both components. Consequently, the following Fourier modes are examined
\begin{equation}\label{eq3.23}
\bar{u}_{j}^{n}=\lambda^{n} e^{i \kappa j h}, \quad h \overline{D u}_{j}^{n}=i \kappa h \mu^{n} e^{i \kappa j h}, \quad \kappa h \in[0,2 \pi], \quad i^{2}=-1
\end{equation}
where $\lambda$ and $\mu$ denote amplification factors. Note that $\lambda\neq\mu$ due to distinct numerical formulae governing the updates of the solution and its derivative. The schemes are expressed in matrix form

\begin{align}
\bar{u}_{j}^{n+1} & =\sum_{m} a_{m}(\nu) \bar{u}_{j+m}^{n}+h \sum_{m} b_{m}(\nu) \overline{D u}{ }_{j+m}^{n},  \\
h \overline{D u}_{j}^{n+1} & =\sum_{m} c_{m}(\nu) \bar{u}_{j+m}^{n}+h \sum_{m} d_{m}(\nu) \overline{D u}_{j+m}^{n}.
\end{align}
Simplify the above equations by inserting $\bar{u}_{j}^{n+1}$ and $\overline{D u}_{j}^{n+1}$ from different schemes in \eqref{eq3.23} to get
\begin{align}
\left[\begin{array}{c}
\bar{u}_{j}^{n+1} \\
h \overline{D u}_{j}^{n+1}
\end{array}\right]=\left[\begin{array}{cc}
\sum_{m} a_{m}(\nu) e^{i m \kappa h} & \sum_{m} b_{m}(\nu) e^{i m \kappa h} \\
\sum_{m}^{m} c_{m}(\nu) e^{i m \kappa h} & \sum_{m} d_{m}(\nu) e^{i m \kappa h}
\end{array}\right]\left[\begin{array}{c}
{\overline{u}}_{j}^{n} \\
h \overline{D u}_{j}
\end{array}\right],
\end{align}
where $a_{m}, b_{m}, c_{m}, d_{m}$ are scheme-dependent coefficients. Hence the spectrum of the matrix is obtained as follows:
\begin{align}\label{eq3.44}
 G=\left[\begin{array}{ll}
\sum_{m} a_{m}(\nu) e^{i m \kappa h} & \sum_{m} b_{m}(\nu) e^{i m \kappa h}  \\
\sum_{m}^{m} c_{m}(\nu) e^{i m \kappa h} & \sum_{m} d_{m}(\nu) e^{i m \kappa h}
\end{array}\right]   
\end{align}

Denote by $\rho_{m}, m=1,2$ eigenvalues of $G$, with $ \Lambda=\operatorname{diag}\left(\rho_{1}, \rho_{2}\right)$. The stability criterion is dictated by the maximal modulus of $\rho_{m}(G)$.
\begin{equation}\label{eq3.60}
\max _{\kappa h \in[0,2 \pi]}\left|\rho_{m}(G)\right| \leq 1+\mathcal{O}(\tau).
\end{equation}

A comparative analysis of stability is conducted for the aforementioned schemes under a uniform temporal discretization framework employing the RK2 and S1O2 methods as follows. 
\subsection{Fully discretized form of LW solver}

Considering conservation laws
\begin{equation}\label{eq3.0}
    \mathbf{W}_t+\nabla \cdot \mathbf{F}(\mathbf{W})=0,
\end{equation}
where $\mathbf{W}$ is a conservative variable and $\mathbf{F}(\mathbf{W})$ is the corresponding flux. The semi-discrete form of a finite volume scheme for equations \eqref{eq3.0} can be written as
\begin{equation}\label{eq3.01}
    \frac{\partial\mathbf{W}_i }{\partial \mathbf{t}} =-\nabla \cdot\mathbf{F}(\mathbf{W})=-\frac{1}{\Delta x}\left({\mathbf{F}_{i+1 / 2}(\mathbf{W})} -{\left.\mathbf{F}_{i-1 / 2}(\mathbf{W})\right)}\right):=\mathcal{L}_{i}(\mathbf{W}),
\end{equation}
where $\mathcal{L}$ is an operator for the spatial derivative of the flux.

The scalar linear advection equation is given as 
\begin{equation}
   u_{t}+c u_{x}=0,
\end{equation}
where $c$ is the constant propagating speed. We assume $c > 0$ here.

In the finite-volume framework, the semi-discretization form could be written as
\begin{equation}
    \bar{u}_{t}=-\frac{1}{\Delta x}\left[F_{i+1 / 2}(t)-F_{i-1 / 2}(t)\right].
\end{equation}

The numerical approximation of the flux $F(t)$, denoted by $\hat{F}(t)$, can be obtained in the finite discretized domain and the above equation could be rewritten as
\begin{equation}\label{eq3.1}
\bar{u}_{t}=-\frac{1}{\Delta x}\left[\hat{F}_{i+1 / 2}(t)-\hat{F}_{i-1 / 2}(t)\right].
\end{equation}
 
Using the definition in \eqref{eq3.01}, then \eqref{eq3.1} could be processed as a time-dependent equation
\begin{equation}
\bar{u}_{t}=\mathcal{L},
\end{equation}
where 
\begin{equation}
\mathcal{L}=-\frac{1}{\Delta x}\left[\hat{F}_{i+1 / 2}(t)-\hat{F}_{i-1 / 2}(t)\right],
\end{equation}
while
\begin{equation}
\mathcal{L}_{t}=-\frac{1}{\Delta x}\left[\hat{F}_{t,i+1 / 2}(t)-\hat{F}_{t,i-1 / 2}(t)\right].
\end{equation}

To reach second-order spatial accuracy, $\hat{u}_{j+\frac{1}{2}}$ is approximated by the upwind scheme, i.e. the cell interface value is updated by
\begin{equation}\label{eq3.7}
 \hat{u}_{j+\frac{1}{2}}= \bar{u}_{j}^{n}+(u_{x})_{j}^{n}\frac{\Delta x}{2}
\end{equation}

\begin{remark}
This section focuses on CGKS, as it is identical to the GRP solver in second-order spatial accuracy.
\end{remark}
\subsubsection{CGKS with RK2}
Considering second-order Runge–Kutta CGKS, the cell averaged derivative in cell $j$ is updated by
\begin{align}\label{eq3.8}
   (u_x)_{j}^{n+1} \Delta x &= \hat{u}_{j+\frac{1}{2}}^{n+1} - \hat{u}_{j-\frac{1}{2}}^{n+1} \nonumber\\
   &=(u_{j+\frac{1}{2}}^{n}+\Delta t(u_{t})_{j+\frac{1}{2}^{n}})-(u_{j-\frac{1}{2}}^{n}+\Delta t(u_{t})_{j-\frac{1}{2}^{n}})\nonumber\\
   &= \bar{u}_{j+\frac{1}{2}}^{n}-\bar{u}_{j-\frac{1}{2}}^{n}+\Delta t((u_{t})_{j+\frac{1}{2}}^{n}-(u_{t})_{j+\frac{1}{2}}^{n})\nonumber\\
   &=(u_x)_{j}^{n} \Delta x+c\frac{\Delta t}{\Delta x}((u_x)_{j-1}^{n} \Delta x-(u_x)_{j}^{n} \Delta x).
\end{align}

Substituting \eqref{eq3.7} and \eqref{eq3.8} into second-order Runge–Kutta temporal discretization
\begin{align}
k_1 &= \mathcal{L}(\bar{u}^n),\label{eq3.9} \\
k_2 &= \mathcal{L}\left(\bar{u}^n + \Delta t \cdot k_1\right), \\
\bar{u}^{n+1} &= \bar{u}^n + \frac{\Delta t}{2} \left(k_1 + k_2\right).\label{eq3.11}
\end{align}

Then the numerical solution at $t^{n+1}$ is obtained by
\begin{align}\label{eq3.12}
    \bar{u}_{j}^{n+1}=&\bar{u}_{j}^{n}+\nu (\bar{u}_{j-1}^{n}-\bar{u}_{j}^{n})+\frac{\nu}{2}\Delta x((u_{x})_{j-1}^{n}-(u_{x})_{j}^{n})+\frac{\nu^{2}}{2}(\bar{u}_{j-2}^{n}-2\bar{u}_{j-1}^{n}+\bar{u}_{j}^{n}) \nonumber\\
    &+\frac{\nu^{2}}{2}\Delta x((u_{x})_{j-2}^{n}-2(u_{x})_{j-1}^{n}+(u_{x})_{j}^{n}),
\end{align}
where $\nu=c\frac{\Delta t}{\Delta x}$. 

Then, the truncation error of the cell at time $t^{n}$ is
\begin{equation}\label{eq3.41}
    T^{n}_{j}=-\frac{(\Delta x)^3}{12}(2\nu^3+\nu)\frac{\partial^3u}{\partial x^3}(x_{j},t_{n})+\frac{(\Delta x)^4}{24}(\nu^4+5\nu^2+\nu )\frac{\partial^4u}{\partial x^4}(x_{j},t_{n})+\mathcal{O}((\Delta x)^5).
\end{equation}
\begin{remark}
 The modified equation is derived via Taylor series expansion. To rigorously quantify the dispersion and dissipation errors, recursive substitution is employed in this and the subsequent subsections: the leading-order approximation of the governing equation itself is used to eliminate high-order time derivatives in the truncation error terms (third order and above).
\end{remark}
\subsubsection{CGKS with spatial-temporal coupled S1O2}
The one-stage second-order temporal discretization is briefly presented for the time-dependent governing equation
$$\frac{\partial u}{\partial t}=\mathcal{L}(u)$$
with the initial data $\left.u(t)\right|_{t=t_{n}}=u^{n}$

 Within the CGKS-S1O2 framework, the general formulation of cell averaged value for one-stage second-order temporal discretization is updated by
\begin{equation}
u^{n+1}=u^{n}+\Delta t \mathcal{L}\left(u^{n}\right)+\frac{1}{2} \Delta t^{2}\frac{\partial}{\partial t} \mathcal{L}\left(u^{n}\right).
\end{equation}
Then the numerical solution at $t^{n+1}$ is obtained by
\begin{align}\label{eq3.14}
    \bar{u}_{j}^{n+1}=&\bar{u}_{j}^{n}+\nu (\bar{u}_{j-1}^{n}-\bar{u}_{j}^{n})+\frac{\nu}{2}\Delta x((u_{x})_{j-1}^{n}-(u_{x})_{j}^{n})+\frac{\nu^{2}}{2}\Delta x((u_{x})_{j}^{n}-(u_{x})_{j-1}^{n}),
\end{align}
where $\nu=c\frac{\Delta t}{\Delta x}$.

Then, the truncation error of the cell at time $t^{n}$ is
\begin{equation}\label{eq3.42}
    T^{n}_{j}=-\frac{(\Delta x)^3}{12}(2\nu^3-3\nu^2+\nu)\frac{\partial^3u}{\partial x^3}(x_{j},t_{n})+\frac{(\Delta x)^4}{24}(\nu^4-2\nu^2+\nu)\frac{\partial^4u}{\partial x^4}(x_{j},t_{n})+\mathcal{O}((\Delta x)^5).   
\end{equation}

\begin{figure}[htbp]
\centering
\begin{minipage}[t]{0.49\textwidth}
\centering
\includegraphics[width=6cm]{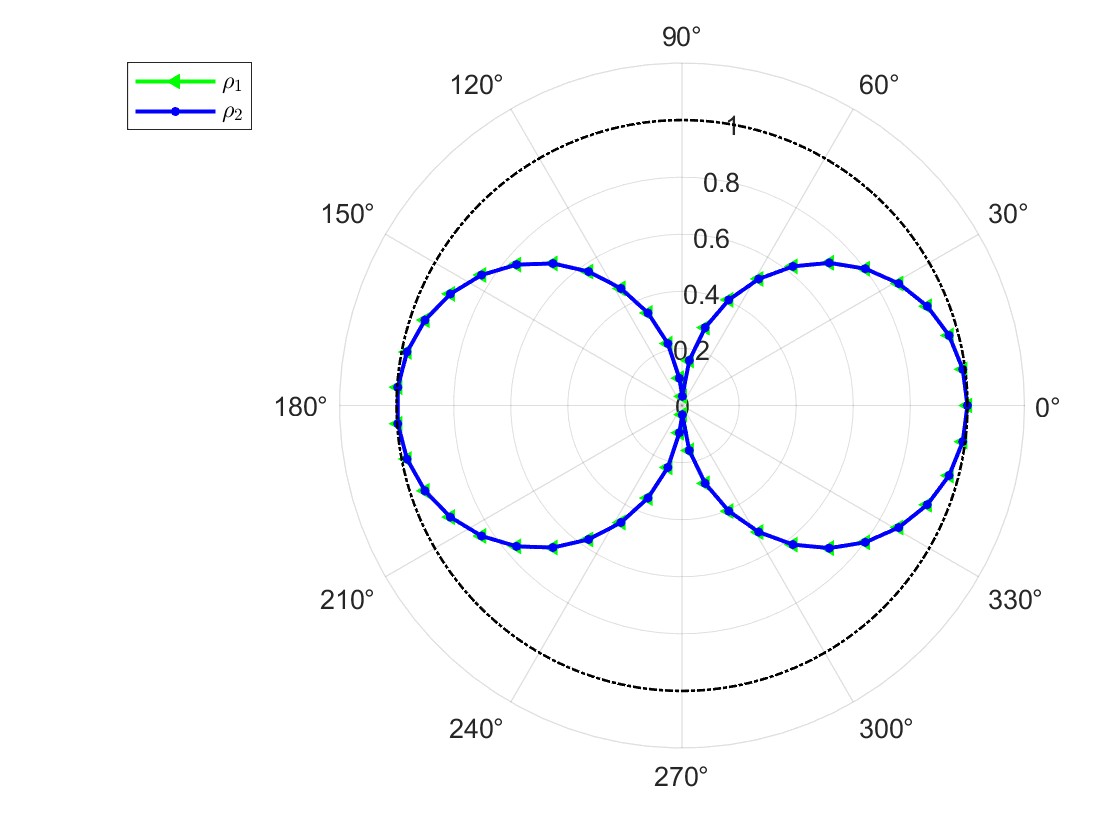}\\
\caption{CGKS-RK-CFL=1.0}\label{fig1}
\end{minipage}
\begin{minipage}[t]{0.49\textwidth}
\centering
\includegraphics[width=6cm]{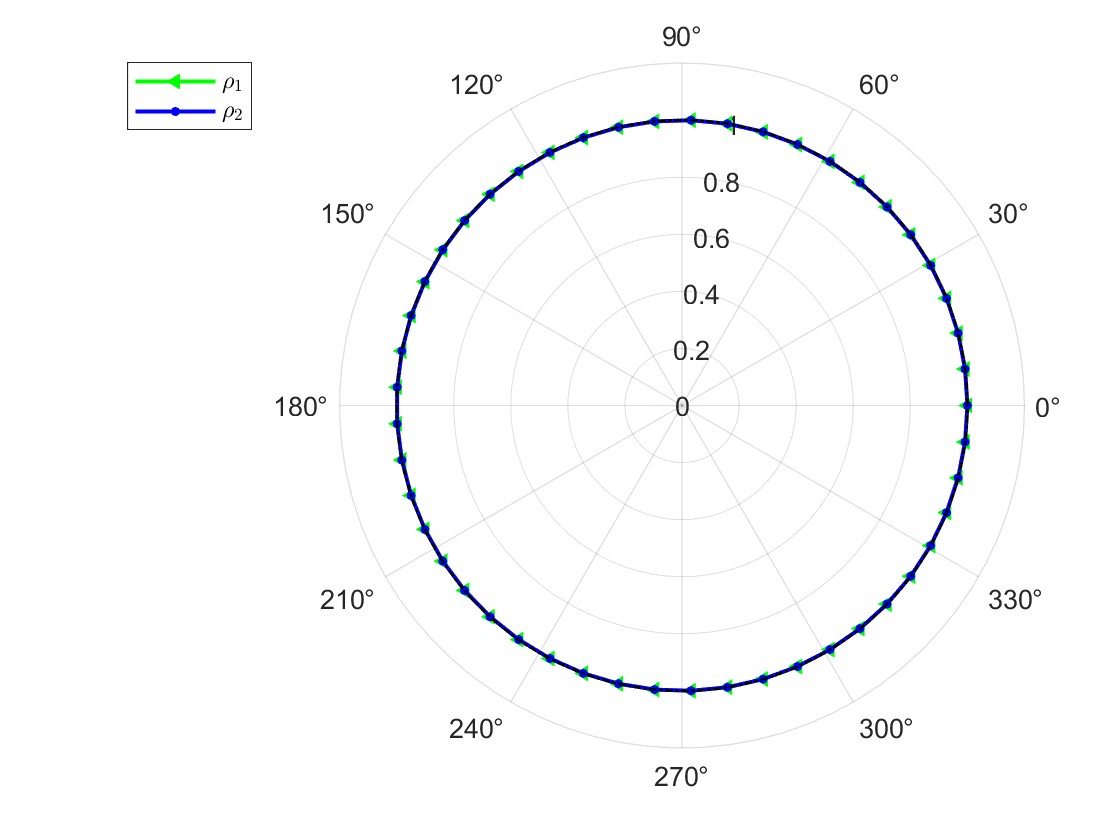}\\
\caption{CGKS-S1O2-CFL=1.0}\label{fig2}
\end{minipage}
\end{figure}

The truncation error \eqref{eq3.41} and \eqref{eq3.42} demonstrate less dispersion and less dissipation within spatial-temporal coupled method than the method of line. Fig.\ref{fig1} and Fig.\ref{fig2} show the eigenvalues $\rho_{m}(G)$ in the complex plane for $\kappa h \in[0,2 \pi]$ for \eqref{eq3.44}. All eigenvalues lie inside the unit circle, indicating the good stability property of the schemes within the different CFL numbers. Moreover, their proximity to the unit circle implies better high-order stability theoretically.

\subsection{Fully discretized form of RK type solver}
To achieve second-order temporal accuracy, the construction of DG and FR methods simplifies to: employing linear polynomial reconstruction within individual grid cells while adopting the Legendre orthonormal basis \cite{CDLWL}.
\subsubsection{DG solver}

(\romannumeral1) \textbf{DG with RK2}
Conventionally, two-stage Runge-Kutta schemes are utilized for $\operatorname{DG}\left(p_{1}\right)$.
The spatial discretization yields a system of ordinary differential equations: 
\begin{equation}
\frac{\mathrm{d}\boldsymbol{u}_{j}}{\mathrm{d}t}=\boldsymbol{M}^{-1}\boldsymbol{R}(\boldsymbol{u}_{j})
\end{equation}
where $\boldsymbol{u}_{j}=(u_{j}^{(0)}(t),u_{j}^{(1)}(t))^{\mathrm{T}}$ represents the vector of unknown degrees of freedom.
Substituting \eqref{eq2.19} into RK2 temporal discretization \eqref{eq3.9}-\eqref{eq3.11}, the cell average value of numerical solution at $t^{n+1}$ is obtained by

\begin{align}\label{eq3.16}  
 \bar{u}_{j}^{n+1}=&\bar{u}_{j}^{n}+\nu (\bar{u}_{j-1}^{n}-\bar{u}_{j}^{n})+\frac{\nu}{2}\Delta x((u_{x})_{j-1}^{n}-(u_{x})_{j}^{n})-\nu^{2}(\bar{u}_{j-2}^{n}-2\bar{u}_{j-1}^{n}+\bar{u}_{j}^{n}) \nonumber\\
    &+\frac{\nu^{2}}{2}\Delta x(2(u_{x})_{j}^{n}-(u_{x})_{j-1}^{n}-(u_{x})_{j-2}^{n}).
\end{align}

Then, the truncation error of the cell at time $t^{n}$ is
\begin{equation}\label{eq3.45}
      T^{n}_{j}=-\frac{(\Delta x)^3}{12}(2\nu^3-3\nu^2+\nu)\frac{\partial^3u}{\partial x^3}(x_{j},t_{n})+\frac{(\Delta x)^4}{24}(\nu^4-4\nu^2+\nu)\frac{\partial^4u}{\partial x^4}(x_{j},t_{n})+\mathcal{O}((\Delta x)^5).   
\end{equation}

(\romannumeral2) \textbf{DG with spatial-temporal coupled S1O2}
Based on the one-stage second-order temporal discretization framework described above, the formulation of DG $\left(p_{1}\right)$ proceeds through the following steps:

Compute the vector at $t=t^{n+1}$
\begin{equation}\label{eq3.21}
    \boldsymbol{u}^{n+1}=\boldsymbol{u}^{n}+\Delta t \mathcal{L}\left(\boldsymbol{u}^{n}\right)+\frac{1}{2} \Delta t^{2} \frac{\partial}{\partial t} \mathcal{L}\left(\boldsymbol{u}^{n}\right).
\end{equation}

Specifically, the $i$-th component of $\mathcal{L}\left(\boldsymbol{u}^{n}\right)$ is
\begin{align}
\mathcal{L}^{(i)}\left(\boldsymbol{u}^{n}\right)= & -\sum_{\Gamma \in \partial \Omega_{e}} \int_{\Gamma} \boldsymbol{F}\left(\widehat{\boldsymbol{u}}_{h}^{n}\left(x, t_{n}\right)\right)  \cdot \boldsymbol{V}_{h}^{i}(x) \mathrm{d} s +\int_{\Omega_{e}} \boldsymbol{F}\left(\boldsymbol{u}_{h}^{n}\left(x, t_{n}\right)\right)\frac{\partial\boldsymbol{V}_{i}(x)}{\partial x} \mathrm{d} \Omega, \quad 0 \leq i \leq N
\end{align}
where the numerical flux function $\widehat{\boldsymbol{F}_{k}}\left(\boldsymbol{u}_{h}\right)$ is rewritten as $\boldsymbol{F}\left(\widehat{\boldsymbol{u}}_{h}^{n}\left(x, t_{n}\right)\right)$ to simplify the derivation of flux Jacobian matrix and $\widehat{\boldsymbol{u}}_{h}^{n}\left(x, t_{n}\right)$ is determined through either approximate or exact Riemann problem.

The temporal derivatives of the residual vector $\frac{\partial}{\partial t} \boldsymbol{L}\left(\boldsymbol{u}^{n}\right)$ constitute the critical component in implementating the one-stage second-order $\operatorname{DG}\left(p_{1}\right)$ method. These derivatives are computed via a component-wise procedure as follows
\begin{align}
\frac{\partial}{\partial t} L^{(i)}\left(\boldsymbol{u}^{n}\right)= & -\sum_{\Gamma \in \partial \Omega_{e}} \int_{\Gamma} \frac{\partial \boldsymbol{F}\left(\widehat{\boldsymbol{u}}_{h}^{n}\left(x, t_{n}\right)\right)}{\partial t} \cdot \boldsymbol{V}_{i}(x) \mathrm{d} s+\int_{\Omega_{e}} \frac{\partial \boldsymbol{F}\left(\boldsymbol{u}_{h}^{n}\left(x, t_{n}\right)\right)}{\partial t}\frac{\partial\boldsymbol{V}_{i}(x)}{\partial x}  \mathrm{d} \Omega, 
\end{align}
where $0 \leq i \leq N $.

Considering the temporal derivatives of the flux function, the Lax-Wendroff procedure could be employed to get 
\begin{align}
 \frac{\partial \boldsymbol{F}\left(\boldsymbol{u}_{h}^{n}\left(x, t_{n}\right)\right)}{\partial t} &= \left.\frac{\partial \boldsymbol{F}\left(\boldsymbol{u}_{h}^{n}\left(x, t_{n}\right)\right)}{\partial \boldsymbol{u}_{h}^{n}\left(x, t_{n}\right)} \frac{\partial \boldsymbol{u}_{h}^{n}(x, t)}{\partial t}\right|_{t=t_{n}} \nonumber\\
 &= -c\left.\frac{\partial \boldsymbol{F}\left(\boldsymbol{u}_{h}^{n}\left(x, t_{n}\right)\right)}{\partial \boldsymbol{u}_{h}^{n}\left(x, t_{n}\right)} \frac{\partial \boldsymbol{u}_{h}^{n}(x, t)}{\partial x}\right|_{t=t_{n}}.\nonumber
\end{align}

Now the solution is updated temporally using the DG solution vector at time $t_{n}$ through the formula \eqref{eq3.21}, thereby obtaining the numerical solution at $t^{n+1}$
\begin{align}\label{eq3.20}
    \bar{u}_{j}^{n+1}=&\bar{u}_{j}^{n}+\nu (\bar{u}_{j-1}^{n}-\bar{u}_{j}^{n})+\frac{\nu}{2}\Delta x((u_{x})_{j-1}^{n}-(u_{x})_{j}^{n}) +\frac{\nu^{2}}{2}\Delta x((u_{x})_{j}^{n}-(u_{x})_{j-1}^{n}).
\end{align}

Then, the truncation error of cell at time $t^{n}$ is
\begin{equation}\label{eq3.46}
     T^{n}_{j}=-\frac{(\Delta x)^3}{12}(2\nu^3-3\nu^2+\nu)\frac{\partial^3u}{\partial x^3}(x_{j},t_{n})+\frac{(\Delta x)^4}{24}(\nu^4-2\nu^2+\nu)\frac{\partial^4u}{\partial x^4}(x_{j},t_{n})+\mathcal{O}((\Delta x)^5).   
\end{equation}
Furthermore, it is rigorously established that the DG method retains a CFL condition of $\frac{1}{2k+1}$ \cite{BS1,BS2}.

\begin{figure}[htbp]
\centering
\begin{minipage}[t]{0.49\textwidth}
\centering
\includegraphics[width=6cm]{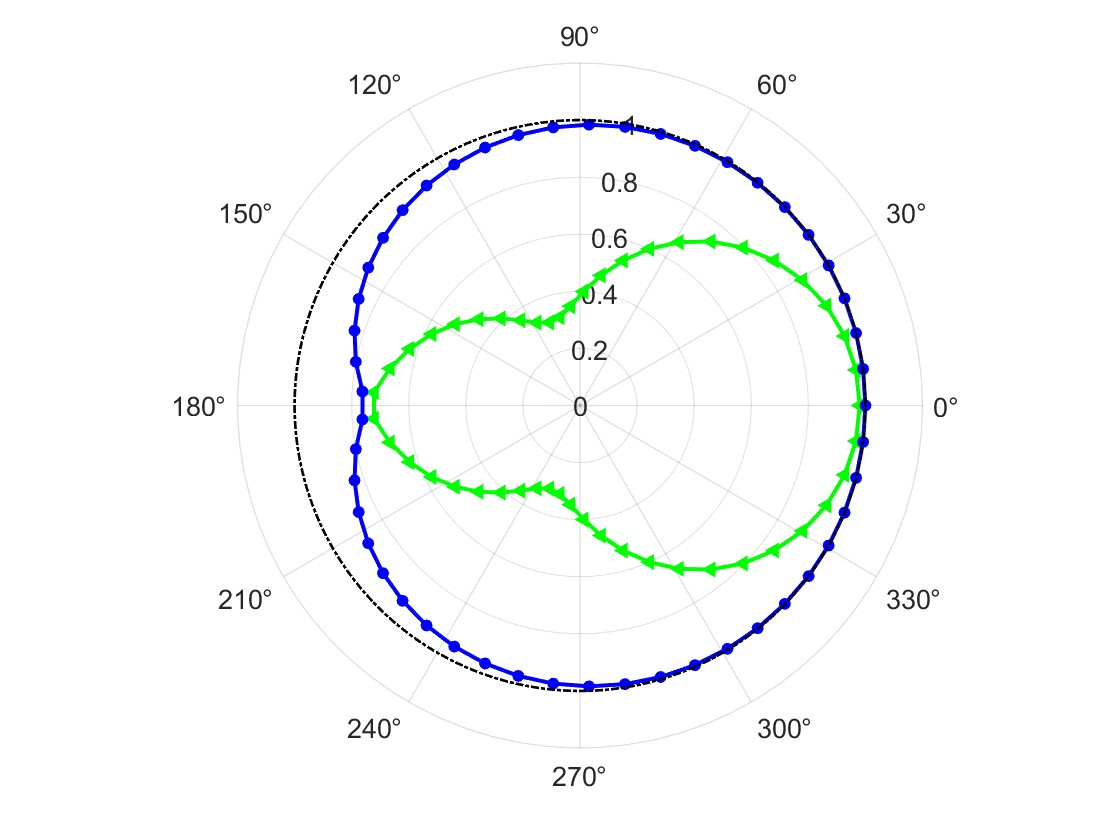}\\
\caption{DG-RK-CFL=0.33}\label{fig3}
\end{minipage}
\begin{minipage}[t]{0.49\textwidth}
\centering
\includegraphics[width=6cm]{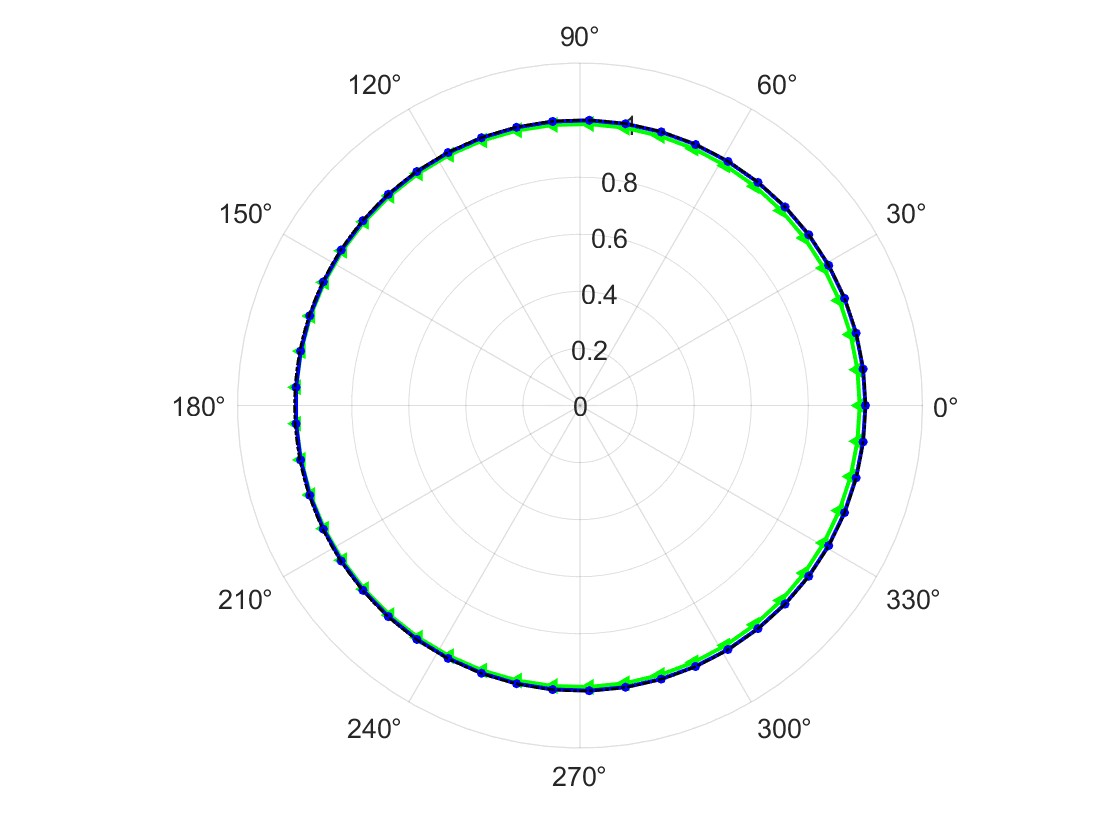}\\
\caption{DG-S1O2-CFL=0.33}\label{fig4}
\end{minipage}
\end{figure}

Here, the truncation error \eqref{eq3.45} and \eqref{eq3.46} demonstrate same dispersion and different dissipation within the two method. The DG within the one-stage second-order  exhibit bigger dissipation while within the RK2 framework exists negative dissipative behavior. Similarly, Fig.\ref{fig3} and Fig.\ref{fig4} exhibit the eigenvalues $\rho_{m}(G)$ in the complex plane for $\kappa h \in[0,2 \pi]$ for \eqref{eq3.44}. All eigenvalues lie inside the unit circle with the CFL numbers limit to $\frac{1}{3}$, where the modulus of eigenvalues lie closer to $1$ for S1O2 compared to RK2.

\subsubsection{FR solver}

Naturally, two-stage Runge-Kutta schemes are utilized for $\operatorname{FR}\left(p_{1}\right)$.
The spatial discretization yields a system of ordinary differential equations with  and  correction functions respectively:
\begin{equation*}
\frac{\mathrm{d} u_{j}^{e}}{\mathrm{~d} t}(t)=-\frac{1}{\Delta x_{e}} \frac{\partial f_{h}}{\partial \xi}(\xi_{j}, t), \quad 0 \leq j \leq N    
\end{equation*}
where $g_{L}(\xi)=\frac{3\xi^2}{4}-\frac{\xi}{2}-\frac{1}{4}$, $g_{R}(\xi)=\frac{3\xi^2}{4}-\frac{\xi}{2}-\frac{1}{4}$ for the Radau correction function and  $g_{L}(\xi)=\frac{3\xi^2}{4}-\frac{\xi}{2}-\frac{1}{4}$, $g_{R}(\xi)=\frac{3\xi^2}{4}-\frac{\xi}{2}-\frac{1}{4}$ for the $g_2$ correction function. As remark \ref{re2.1} motioned, different correction function demonstrate distinct properties and CFL limit, which will be shown in section \ref{sec3.4}. 

Substituting \eqref{eq2.21}-\eqref{eq2.22} into RK2 temporal discretization \eqref{eq2.23} with the Radau correction function to obtain the cell average value $\bar{u}_{j}^{n+1}$ by
\begin{align}\label{eq3.30}  
 \bar{u}_{j}^{n+1}=&\bar{u}_{j}^{n}+\nu (\bar{u}_{j-1}^{n}-\bar{u}_{j}^{n})+\frac{\nu}{2}\Delta x((u_{x})_{j-1}^{n}-(u_{x})_{j}^{n})-\nu^{2}(\bar{u}_{j-2}^{n}-2\bar{u}_{j-1}^{n}+\bar{u}_{j}^{n}) \nonumber\\
    &+\frac{\nu^{2}}{2}\Delta x(2(u_{x})_{j}^{n}-(u_{x})_{j-1}^{n}-(u_{x})_{j-2}^{n}),
\end{align}
which is identical to \eqref{eq3.16}. Then, the truncation error of the cell at time $t^{n}$ is
\begin{equation}\label{eq3.47}
      T^{n}_{j}=-\frac{(\Delta x)^3}{12}(2\nu^3-3\nu^2+\nu)\frac{\partial^3u}{\partial x^3}(x_{j},t_{n})+\frac{(\Delta x)^4}{24}(\nu^4-4\nu^2+\nu)\frac{\partial^4u}{\partial x^4}(x_{j},t_{n})+\mathcal{O}((\Delta x)^5).   
\end{equation}

Likewise, substitute \eqref{eq2.21}-\eqref{eq2.22} into RK2 temporal discretization \eqref{eq2.23} with $g_2$ correction function, the value of $\bar{u}_{j}^{n+1}$ is obtained by
\begin{align}\label{eq3.31}  
 \bar{u}_{j}^{n+1}=&\bar{u}_{j}^{n}+\nu (\bar{u}_{j-1}^{n}-\bar{u}_{j}^{n})+\frac{\nu}{2}\Delta x((u_{x})_{j-1}^{n}-(u_{x})_{j}^{n})+\frac{3\nu^{2}}{2}(\bar{u}_{j-2}^{n}-2\bar{u}_{j-1}^{n}+\bar{u}_{j}^{n}) \nonumber\\
    &+\frac{\nu^{2}}{4}\Delta x(3(u_{x})_{j-2}^{n}-2(u_{x})_{j-1}^{n}-(u_{x})_{j}^{n}).
\end{align}
And the truncation error of the cell at time $t^{n}$ is
\begin{equation}\label{eq3.48}
 T^{n}_{j}=-\frac{(\Delta x)^3}{12}(2\nu^3-3\nu^2+\nu)\frac{\partial^3u}{\partial x^3}(x_{j},t_{n})+\frac{(\Delta x)^4}{24}(\nu^4+\nu^2+\nu)\frac{\partial^4u}{\partial x^4}(x_{j},t_{n})+\mathcal{O}((\Delta x)^5).
\end{equation}

\begin{figure}[htbp]
\centering
\begin{minipage}[t]{0.49\textwidth}
\centering
\includegraphics[width=6cm]{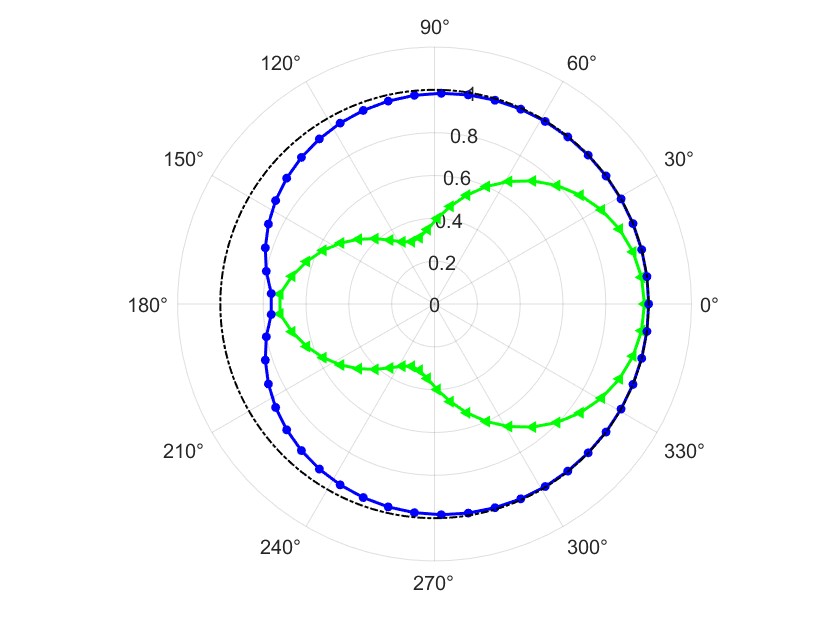}\\
\caption{FR-RK-CFL=0.33 with $g_{Radau}$ correction function}\label{fig8}
\end{minipage}
\begin{minipage}[t]{0.49\textwidth}
\centering
\includegraphics[width=6cm]{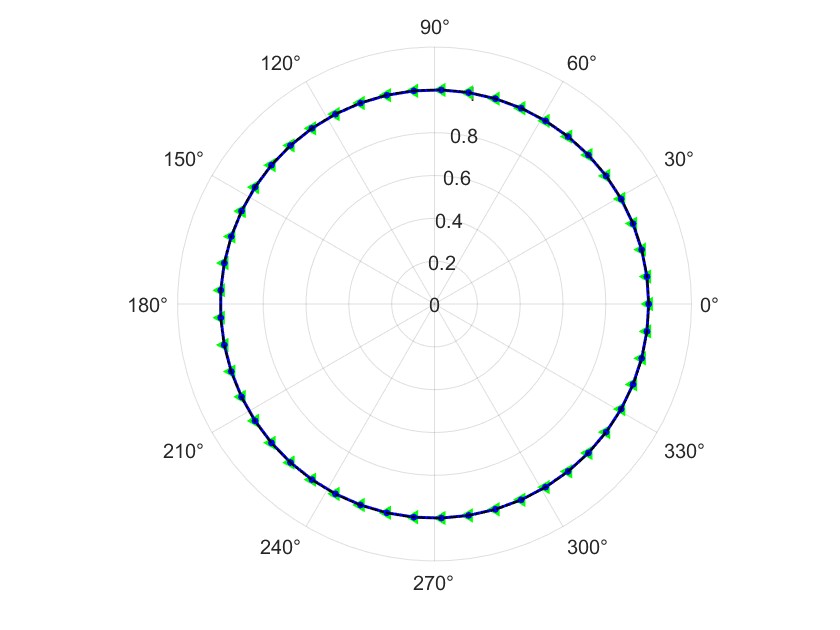}\\
\caption{FR-RK-CFL=1.0 with $g_{2}$ correction function}\label{fig9}
\end{minipage}
\end{figure}

The truncation error \eqref{eq3.47} and \eqref{eq3.48} demonstrate same dispersion and different dissipation within the two correction functions. Within the RK2 framework, the FR method with $g_{2}$ correction function exhibit bigger dissipation while within the $g_{Radau}$ correction function exists negative dissipative behavior. Similarly, Fig.\ref{fig8} and Fig.\ref{fig9} exhibit the eigenvalues $\rho_{m}(G)$ in the complex plane for $\kappa h \in[0,2 \pi]$ for \eqref{eq3.44}. The FR scheme using the $g_{Radau}$ correction function is constrained by the CFL stability limit of $1/3$. Conversely, using the $g_{2}$ correction function increases the permissible CFL number to 1.0. The spectrum of scheme $g_{2}$ turns out to be identical to the boundary of the stability region of the explicit RK2 time stepping method. Besides, the spatial-temporal coupled method is inherently restricted to high-order implementations by its design.

\subsection{Difference of eigenvalues}
 Within the spatial-temporal coupled framework, CGKS and DG attain comparable dispersion errors. However, as shown in Fig.\ref{fig2} and Fig.\ref{fig4}, eigenvalue distributions in the complex plane indicates that CGKS admits a larger CFL number, resulting from the inherent structural distinctions between the numerical schemes. Despite the same dispersion and dissipation, the eigenvalues of $G$ for compact gas kinetic scheme within the S1O2 and RK2 framework could technically be triggered. For S1O2, the eigenvalues of $G$ are
\begin{equation}\label{eq3.61}
   \rho_{1} =\rho_{2} =1 - \nu (1 - e^{-i\theta})
\end{equation}
where $\theta=\kappa h$,$\theta \in[0,2 \pi]$. Substituting the \eqref{eq3.61} into the \eqref{eq3.60} to get
\begin{equation*}
    (1+\nu \cos{\theta}-\nu)^{2}+\nu^{2}\sin{\theta}^{2}\leq 1,
\end{equation*}
which yields the following necessary and sufficient stability condition 
\begin{equation*}
    0\leq \nu\leq 1.
\end{equation*}

For RK2, the eigenvalues of $G$ are
\begin{equation}\label{eq3.62}
    \rho_{1}=\rho_{2}=1+\nu (e^{-i\theta} - 1)+\frac{\nu^2}{2}(e^{-i\theta} - 1)^2.
\end{equation}
Substituting the \eqref{eq3.62} into the \eqref{eq3.60} to get the following inequalities
\begin{align}\label{eq4.1}
  \left\{\begin{array}{ll}
  0<\nu,\quad\theta=0,\\
 0< \nu \leq Root[\Theta(y,\theta),1],\quad0<\theta<\pi \\
 0\leq\nu\leq1,\quad\theta=\pi\\
 0< \nu \leq Root[\Theta(y,\theta),1],\quad\pi<\theta<2\pi\\
 0<y,\quad x=2\pi
 \end{array}\right.
\end{align}
where $\Theta=-1-tan^2(\frac{\theta}{2})+2\nu tan^2(\frac{\theta}{2})-2\nu^2tan^2(\frac{\theta}{2})+\nu^3tan^2(\frac{\theta}{2})$,  and $Root[\Theta(y,\theta)]$ represent the minimum real root of equation $\Theta(y,\theta)=0$, which could be simplified as
\begin{equation*}
    \nu^3-2\nu^2+2\nu-\left(1+\frac{1}{tan^2(\frac{\theta}{2})}\right)=0.
\end{equation*}

When $x\rightarrow \pi^{+}$, one get $Root[\Theta(y,\theta)]=1$. Hence, the minimum value of $Root[\Theta(y,\theta)]$ equals to $1$ when $\theta$ varies in $[0,2\pi]$. Therefore, 
\begin{equation*}
    0\leq \nu\leq 1
\end{equation*}
is a sufficient but not necessary condition for stability of CGKS within the RK2 framework. Such distinct of eigenvalues between two schemes stem from the evolution of the spatial derivative average of
the cell.

Numerical errors for DG and CGKS under varying CFL condition in both $L_1$ and $L_2$ norms are summarized in the tables below, computed using a uniform mesh of 640 cells.
\begin{table}[htbp]
\caption{Numerical error for velocity perturbation by the CGKS within different CFL number}\label{tab1}
    \begin{center}
    \begin{tabular}{ccccc}
    \hline
        CFL & $L^1$ error & $L^2$ error\\
        \hline
        1.0000 & 8.22204E-06 & 3.60984E-07 \\
        1.0100 & 8.5175E-04 & 3.7396E-05 \\
        1.1000 & 5.4683E+13 & 2.9788E+12\\
        \hline
    \end{tabular}
    \end{center}
\end{table}

\begin{table}[htbp]
\caption{Numerical error for velocity perturbation by the DG within different CFL number}\label{tab2}
\begin{center}
    \begin{tabular}{ccccc}
    \hline
        CFL & $L^1$ error & $L^2$ error\\
        \hline
        0.3333 & 6.08805E-05 & 2.67292E-06 \\
        0.3340 & 1.68979E-03 & 7.41903E-05 \\
        0.3400 & 5.59466E+09 & 2.4563E+08  \\
        \hline
    \end{tabular}
    \end{center}
\end{table}
As evidenced in \cref{tab1} and \cref{tab2}, the CGKS exhibits exponential error growth as its CFL number incrementally rises from $1.0000$ to $1.0100$, with errors reaching magnitudes of $10^{13}$ at CFL = 1.1000. Concurrently, the DG method demonstrates analogous instability: errors surge exponentially when its CFL number increases from $0.3333$ to $0.3340$, attaining magnitudes of $10^{9}$ at CFL = 0.34.

\Cref{tab3} summarizes the aforementioned differences in CFL limit, dispersion and dissipation, with the CGKS with S1O2 serving as the benchmark. Here, the plus sign denotes a larger quantity, and the minus sign a smaller one.
\begin{table}[htbp]
\footnotesize
\caption{Contrast for RK and LW type methods}\label{tab3}
\begin{center}
    \begin{tabular}{ccccc}
    \hline
       Different schemes & CFL Limit & Dispersion & Dissipation & Spatial-temporal Coupled \\
        \hline
        CGKS with RK2&1.0 & + & + &$\times$\\
       CGKS with S1O2 & 1.0 & 0 &0 & $\checkmark$ \\
        DG with RK2& $\frac{1}{3}$ & 0 & - & $\times$ \\
        DG with S1O2& $\frac{1}{3}$ & 0 & 0 & $\checkmark$ \\
        \hline
    \end{tabular}
    \end{center}
\end{table}

\subsection{Dissipation analysis for high-order scheme}
\subsubsection{Two-stage fourth-order temporal discretization}

 For higher-order temporal discretization, the two-stage fourth-order temporal discretization is concisely summarized below. Considering conservation laws, the semi-discrete finite volume scheme is written as
\begin{equation}\label{eq3.28}
 \frac{\partial u}{\partial t}=\mathcal{L}(u) 
\end{equation}
with the initial dat
\begin{equation}
\left.u(t)\right|_{t=t_n}=u^n,   
\end{equation}
where $\mathcal{L}(u)$ is an operator for spatial derivatives.

The generalized two-stage fourth-order temporal discretization framework is given as follows.

Stage 1. Predict the intermediate stage
\begin{equation}
u^{(1)}  =u^n+\frac{1}{2} \Delta t \mathcal{L}\left(u^n\right)+\frac{1}{8} \Delta t^2 \frac{\partial}{\partial t} \mathcal{L}\left(u^n\right),
\end{equation}
\begin{equation}
\frac{\partial}{\partial t} \mathcal{L}\left(u^n\right) =\frac{\partial}{\partial u} \mathcal{L}\left(u^n\right) \mathcal{L}\left(u^n\right)
\end{equation}  
where the second equation derived from applying the chain rule to \eqref{eq3.28}.

Stage 2. Update the solution using the formula
\begin{equation}
  u^{n+1}=u^n+\Delta t \mathcal{L}\left(u^n\right)+\frac{1}{6} \Delta t^2\left(\frac{\partial}{\partial t} \mathcal{L}\left(u^n\right)+2 \frac{\partial}{\partial t} \mathcal{L}\left(u^{(1)}\right)\right) .  
\end{equation}

\subsubsection{CGKS-S2O4}
Within a one-dimensional Hermite WENO framework, the reconstruction of the left interface value $W_{i+1/2}^L$ at $x_{i+1/2}$ employs 
The point value at the cell interface $x_{i+1 / 2}$ is thereby expressed as
\begin{equation*}
  W_{i+1/2}^l=-\frac{23}{120} \bar{W}_{i-1}+\frac{19}{30} \bar{W}_{i}+\frac{67}{120} \bar{W}_{i+1}-\Delta x\left(\frac{3}{40}\left(\bar{W}_{x}\right)_{i-1}+\frac{7}{40}\left(\bar{W}_{x}\right)_{i+1}\right),  
\end{equation*}
where the derivatives of non-equilibrium components $\left(W_{x}\right)_{i+1 / 2}^{e}$ and $\left(W_{x x}\right)_{i+1 / 2}^{e}$ are obtained as follows:
\begin{equation*}
\left(W_{x}\right)_{i+1 / 2}^{e}=\left[-\frac{1}{12}\left(\bar{W}_{i+2}-\bar{W}_{i-1}\right)+\frac{5}{4}\left(\bar{W}_{i+1}-\bar{W}_{i}\right)\right] / \Delta x,
\end{equation*}
\begin{equation*}
\left(\bar{W}_{x x}\right)_{i+1 / 2}^{e}=\left[-\frac{1}{8}\left(\bar{W}_{i+2}+\bar{W}_{i-1}\right)+\frac{31}{8}\left(\bar{W}_{i+1}+\bar{W}_{i}\right)-\frac{15}{2} W_{i+1 / 2}^{e}\right] / \Delta x^{2} .
\end{equation*}

For the CGKS, the cell-averaged values and their derivatives are updated simultaneously to reach the same spatial-temporal accuracy. By introducing an intermediate state at $t_{*}=t_{n}+K \Delta t$,

\begin{equation}\label{eq3.33}
f^{*}=f^{n}+K \Delta t f_{t}^{n}+\frac{1}{2} K^{2} \Delta t^{2} f_{t t}^{n} 
\end{equation}
the state $f^{n+1}$ is updated with the following formula
\begin{equation}\label{eq3.34}
f^{n+1}=f^{n}+\Delta t\left(M_{0} f_{t}^{n}+M_{1} f_{t}^{*}\right)+\frac{1}{2} \Delta t^{2}\left(N_{0} f_{t t}^{n}+N_{1} f_{t t}^{*}\right)
\end{equation}
yielding fourth-order accuracy with the coefficients specified below
\begin{equation}
K=\frac{1}{2}, M_{0}=1, M_{1}=0, N_{0}=\frac{1}{3}, N_{1}=\frac{2}{3}. 
\end{equation}

Consequently, the dissipative behavior of CGKS with two-stage fourth-order temporal discretization is illustrated in Fig. \ref{fig6} below, maintaining numerical stability at a CFL condition of 0.56, whereas the DG method  exhibits a reduced CFL limit of 0.11 achieving equivalent temporal accuracy.
\begin{figure}[htbp]
\centering
\includegraphics[width=7cm]{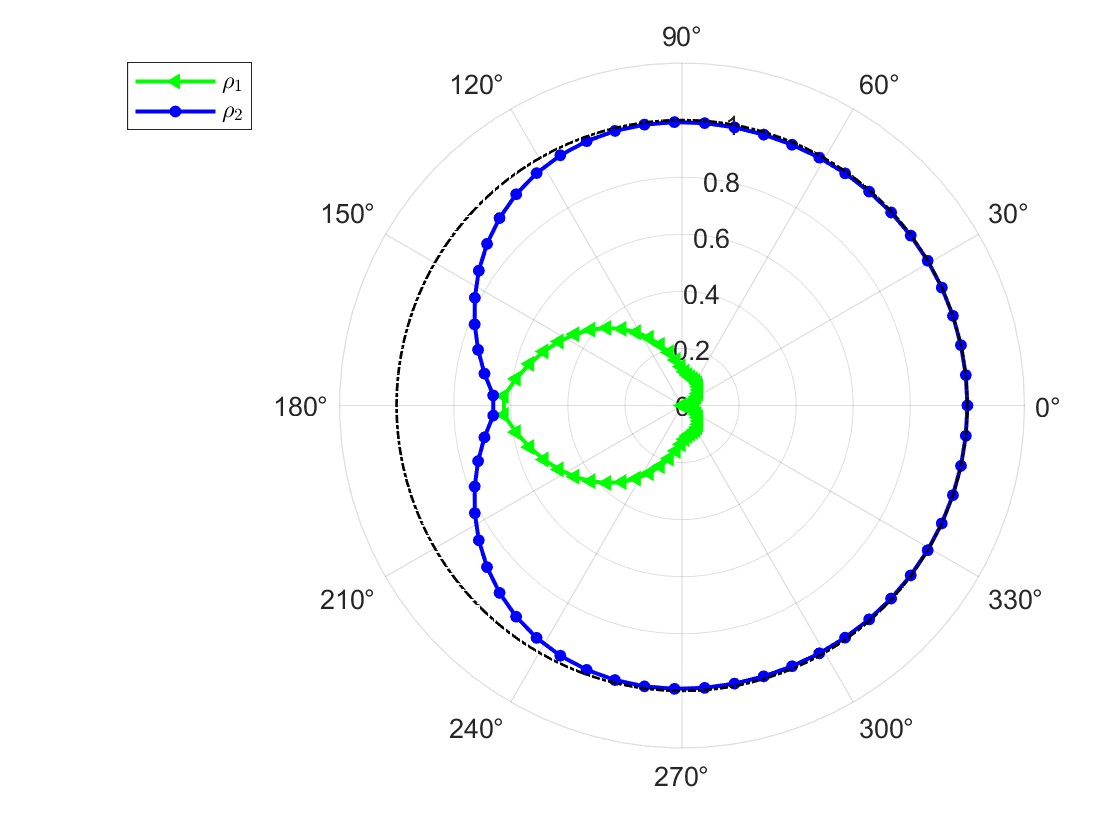}\\
\caption{CGKS-S2O4}\label{fig6}
\end{figure}
\begin{remark}
Based on the definition of the intermediate state, the expansion in \eqref{eq3.33}-\eqref{eq3.34} are transformed as follows
\begin{align}
f^{n+1}-f^{n} & =\Delta t\left(M_{0}+M_{1}\right) f_{t}^{n}+\frac{\Delta t^{2}}{2}\left(N_{0}+N_{1}+2 M_{1} K\right) f_{t t}^{n} \nonumber\\
& +\frac{\Delta t^{3}}{2}\left(M_{1} K^{2}+N_{1} K\right) f_{t t t}^{n}+\frac{\Delta t^{4}}{4} N_{1} K^{2} f_{t t t t}^{n}+\mathcal{O}\left(\Delta t^{5}\right)
\end{align}

To achieve fourth-order temporal accuracy for the interface value at 
$t^{n+1}$, the Taylor expansion
\begin{equation}\label{eq4.7}
   f^{n+1}=f^{n}+\Delta t f_{t}^{n}+\frac{\Delta t^{2}}{2} f_{t t}^{n}+\frac{\Delta t^{3}}{6} f_{t t t}^{n}+\frac{\Delta t^{4}}{24} f_{t t t t}^{n}+\mathcal{O}\left(\Delta t^{5}\right) 
\end{equation}
 uniquely determines the coefficients. Consequently, the macroscopic variables $W_{i+1 / 2}^{n+1}$ at the cell interface are derived by taking moments of $f^{n+1}$ and these interface values serve as initial conditions for the reconstruction at the next time step.
\end{remark}

\section{Discussions}\label{sec4}
 This study is motivated by the contrasting stability limits observed in high-order numerical schemes: while the DG method usually requires a CFL constraint of 0.11 for fourth-order accuracy, the CGKS remains stable at a CFL number of 0.5 \cite{JPSX}. This disparity raises a fundamental question: why does CGKS exhibit superior stability under identical spatial stencil widths and degrees of freedom? We attribute this difference to two principal factors: the inherent differences in flux reconstruction strategies and the role of spatial-temporal coupled property in the time-marching scheme. Consequently, we categorize the methods into two families: the Runge–Kutta type spatial-temporal decoupled method and the Lax–Wendroff (LW) type spatial-temporal coupled methods. Through a comprehensive linear stability analysis within the frameworks established in Sections \ref{sec2}–\ref{sec3}, we address this question from the following three perspectives.
 
 First, it is observed that under the same S1O2 time marching method, the formula for evolving cell-averaged value for both spatial-temporal coupled method and decoupled method are identical. What, then, causes the CFL limit for DG to be 1/3, while that for CGKS is 1? The two-moment formulation reveals that the difference stems from their distinct evolution strategies for the spatial derivative of cell averages. Both GKS and GRP schemes derive their cell-averaged value from exact mathematical-physics models by evolving the solution and its spatial derivative averages simultaneously. This further explains why, in Fig.\ref{fig9}, applying the $g_2$ correction function in flux reconstruction coincidentally modifies the evolution of derivative averages and consequently improves the CFL limit.

 Second, the RK scheme introduces numerical artifacts—named here as "Spatial-temporal incompatibility"—due to its decoupling of spatial and temporal discretizations. A comparative truncation error analysis shows that within the DG framework, both temporal integration methods yield identical dispersion errors; however, the eigenvalues of the spatial-temporal coupled method lie closer to the unit circle than those of the decoupled method. Under the CGKS/GRP framework, spatial-temporal coupled methods further reduce dispersion error compared to the decoupled method. This analysis demonstrates that second-order RK-based temporal discretization produces greater dispersion in spatial-temporal coupled type schemes and negative dissipation in decoupled type schemes. In contrast, the LW-based temporal discretization improves stability on both aspects. By unifying temporal evolution with spatial reconstruction, the LW-type approach maintains physical consistency inherent to hyperbolic systems.

 Finally, as shown in \cref{fig1}–\cref{fig4}, LW-type methods enhance computational efficiency substantially. They permit higher CFL numbers, which—combined with reduced time steps (by half) and more compact stencils—lowers operational costs. The multi-stage nature of RK schemes necessitates expensive limiters and frequent data exchange in parallel environments. In contrast, LW-type methods exploit their spatial-temporal compact design for more efficient temporal evolution. This enables CGKS with S2O4 time integration to maintain stability at a CFL number of 0.56, whose corresponding eigenvalues in the complex plane are shown in \cref{fig6}.

 The previous discussion has been restricted to the linear stability of second-order schemes, comparing the two families of methods via a posteriori evaluation. From a broader perspective, the systematic design of high-order schemes ($\geq$ 3rd-order) that retain stability at large time steps remains an open challenge, lacking a comprehensive theoretical or automated tool. For extending such stability analysis to higher-order and nonlinear problems (e.g., the Navier-Stokes equations), future work must utilize the dissipative matrix formalism and rigorous mathematical analysis—a promising direction for future research in CFD algorithm design.

\section*{Acknowledgments}
The authors would like to thank Dr. Li. Ang for his generous help in two-moment
schemes analysis. This work was funded by the Funding of National Key Laboratory of Computational Physics, the National Natural Science Foundation of China (Grant No. 12302378, 12172316, and 92371107) and the Natural Science Basic Research Plan in Shaanxi Province of China (No. 2025SYS-SYSZD-070).


\begin{thebibliography}{99}
\bibitem{BF}
M. Ben-Artzi, J. Falcovitz, {Generalized Riemann Problems in Computational Fluid Dynamics}, Cambridge
University Press, Cambridge, 2003.

\bibitem{BS1}
B. Cockburn, C.W. Shu, {TVB Runge–Kutta local projection discontinuous Galerkin finite element method for conservation laws II: general framework}, Math. Comput. 52 (1989) 411–435.

\bibitem{BS2} 
B. Cockburn, C.W. Shu, {The Runge–Kutta discontinuous Galerkin method for conservation laws V: multidimensional systems}, J. Comput. Phys. 141 (1998) 199–224.

\bibitem{DFTB} 
M. Dumbser, F. Fambri, M. Tavelli, M. Bader, T. Weinzierl, {Efficient implementation of ADER discontinuous Galerkin schemes for a scalable hyperbolic PDE engine}, Axioms 7 (2018) 63.

\bibitem{CDLWL}
J. Cheng, Z.F. Du, X. Lei, Y. Wang, J.Q. Li,
{A two-stage fourth-order discontinuous Galerkin method based on the GRP solver for the compressible euler equations},
Computers $\&$ Fluids, 181, (2019) 248-258.

\bibitem{HW}
E. Hairer, G. Wanner, {Multistep multistage multiderivative methods for ordinary differential equations}, Computing 11 (1973) 287–303.



\bibitem{Huynh}
H.T. Huynh, A Flux Reconstruction Approach to High-Order Schemes Including Discontinuous Galerkin Methods, AIAA, Miami, FL, June 2007.

\bibitem{JPSX}
X. Ji, L. Pan, W. Shyy, K. Xu, {A compact fourth-order gas-kinetic scheme for the Euler and Navier-Stokes equations}, J. Comput. Phys. 372 (2018) 446–472.

\bibitem{JS}
C.-W. Shu, High order weighted essentially non-oscillatory schemes for convection dominated problems, SIAM Rev. 51 (2009) 82–126.

\bibitem{LD}
J. Li, Z. Du, {A two-stage fourth order time-accurate discretization for Lax–Wendroff type flow solvers I. Hyperbolic conservation laws}, SIAM J. Sci. Comput. 38 (5) (2016) A3046–A3069.
 
\bibitem{LL}
A. Li, J. Li, {Lax-Wendroff solvers-based Hermite reconstruction for hyperbolic problems}, Applied mathematics and computation, 2023.

\bibitem{MZJ}
J. Mu, H. Zhang, Z. Ji, Y. Zhang, G. Chen, K. Xu, https://arxiv.org/abs/2508.13705.

\bibitem{PXL}
L. Pan, K. Xu, Q. Li, J. Li, {An efficient and accurate two-stage fourth-order gas-kinetic scheme for the Navier–Stokes equations}, J. Comput. Phys. 326 
(2016) 197–221.

\bibitem{PJCWX}
L. Pan, J.X. Cheng, S.H. Wang, K. Xu, {A two-stage fourth-order gas-kinetic scheme for compressible multicomponent flows}, Commun. Comput. Phys. 22 (2017) 1123–1149.

\bibitem{QS1}
J.X. Qiu, C.-W. Shu, {Hermite WENO schemes and their application as limiters for Runge–Kutta discontinuous Galerkin method: one-dimensional case}, J. Comput. Phys. 193 (2004) 115–135.

\bibitem{TT}
E. F. Toro, V. A. Titarev, {Derivative Riemann solvers for systems of conservation lawsand ADER methods}, J. Comput. Phys. 212 (2006), pp. 150–165.

\bibitem{TT2}
V.A. Titarev, E.F. Toro, {ADER schemes for three-dimensional non-linear hyperbolic systems}, J. Comput. Phys. 204 (2) (2005)
 715–736.
 
\bibitem{V} 
B. van Leer, {Towards the ultimate conservative difference scheme V. A second order sequel to Godunov’s method}, J. Comput. Phys. 32 (1979) 101–136.

\bibitem{VC} 
P.E. Vincent, P. Castonguay, A. Jameson, {A new class of high-order energy stable flux reconstruction schemes}, J. Sci. Comput. 47 (2011) 50–72.

\bibitem{XJX}
Q. Xie, X. Ji, Z.H. Qiu, C.L. Liang, K. Xu, {High-Order Spectral Difference Gas-Kinetic Schemes for Euler and Navier-Stokes Equations}. East Asian Journal on Applied Mathematics, 13(3), 499-523. 

\bibitem{X2} 
K. Xu, {A gas-kinetic BGK scheme for the Navier–Stokes equations and its connection with artificial dissipation and Godunov method}, J. Comput. Phys. 171 (2001) 289–335.

\bibitem{XH}
K. Xu, J.C. Huang, {A unified gas-kinetic scheme for continuum and rarefied flows}, J. Comput. Phys. 229 (2010) 7747–7764.
 
\end{thebibliography}
\end{document}